\documentclass[conference,letterpaper]{IEEEtran}

\usepackage[utf8]{inputenc}
\usepackage[T1]{fontenc}
\usepackage{url}
\usepackage{ifthen}
\usepackage{cite}
\usepackage[cmex10]{amsmath} 

\usepackage{comment}
\usepackage{amsthm}
\usepackage{amsmath,amssymb,amsfonts}
\usepackage{graphicx}
\usepackage{subfig}
\usepackage{algpseudocode}
\usepackage{algorithm}

\newtheorem{theorem}{Theorem}[section]

\newcommand{\paren}[1]{\left( #1 \right)}
\newcommand{\cbrace}[1]{\left\{#1\right\}}
\newcommand{\sbrace}[1]{\left[#1\right]}
\newcommand{\abs}[1]{\left| #1\right|}

\usepackage{mathtools}

\newcommand{\cC}{{\cal C}}

\newcommand{\cE}{{\cal E}}

\newcommand{\cN}{{\cal N}}

\newcommand{\vc}{\vec{c}}

\newcommand{\vu}{\vec{u}}

\newcommand{\vx}{\vec{x}}
\newcommand{\vy}{\vec{y}}
\newcommand{\vz}{\vec{z}}

\newcommand{\vX}{\vec{X}}
\newcommand{\vY}{\vec{Y}}
\newcommand{\vZ}{\vec{Z}}

\newcommand{\vchat}{\hat{\vc}}
\newcommand{\vxhat}{\hat{\vx}}
\newcommand{\vyhat}{\hat{\vy}}
\newcommand{\vzhat}{\hat{\vz}}

\newcommand{\vZhat}{\widehat{\vZ}}

\newcommand{\gf}{\mathbb{F}}

\newcommand{\gftwo}{\gf_{2}}

\newcommand{\mutualinf}[2]{I\paren{#1;#2}}

\usepackage{xcolor}

\begin{document}
\title{Noise Recycling}



 \author{%
   \IEEEauthorblockN{Alejandro Cohen\IEEEauthorrefmark{1},
                     Amit Solomon\IEEEauthorrefmark{1},
                     Ken R. Duffy\IEEEauthorrefmark{2},
                     and Muriel M\'edard\IEEEauthorrefmark{1}}
   \IEEEauthorblockA{\IEEEauthorrefmark{1}%
                     \textit{RLE, MIT}
                    Cambridge, MA 02139, USA,
                    \{cohenale,amitsol,medard\}@mit.edu}

   \IEEEauthorblockA{\IEEEauthorrefmark{2}%
                     \textit{Hamilton Institute}
                    {Maynooth University, Ireland}, ken.duffy@mu.ie}
 }

\maketitle


\begin{abstract}
We introduce Noise Recycling, a method that enhances decoding
performance of channels subject to correlated noise without joint
decoding. The method can be used with any combination of codes,
code-rates and decoding techniques. In the approach, a continuous
realization of noise is estimated from a lead channel by subtracting
its decoded output from its received signal. This estimate is then
used to improve the accuracy of decoding of an orthogonal channel
that is experiencing correlated noise. In this design, channels aid
each other only through the provision of noise estimates post-decoding.
In a Gauss-Markov model of correlated noise, we constructively
establish that noise recycling employing a simple successive order
enables higher rates than not recycling noise. Simulations illustrate
noise recycling can be employed with any code and decoder, and that
noise recycling shows Block Error Rate (BLER) benefits when applying the
same predetermined order as used to enhance the rate region. Finally, for short codes we establish that an additional BLER improvement
is possible through noise recycling with racing, where the lead channel
is not pre-determined, but is chosen on the fly based on which decoder
completes first.
\end{abstract}


\section{Introduction}
The use of orthogonal channels is commonplace in applications from wired to wireless channels. Examples include the wide-spread use of
orthogonal frequency division multiplexing (OFDM) \cite{NP00,
Yang05}, and of orthogonal schemes in multiple access, such frequency
division multiplexing access (FDMA), time-division multiple access (TDMA),
or orthogonal code-division multiple access (CDMA), see for instance
\cite{Gal85}. In OFDM or FDMA, channels separated by less than a
coherence band will experience correlated noise. Joint decoding of
all orthogonal channels can, in theory, make use of such correlation
to improve performance, but it is challenging to implement efficiently
in practice and, indeed, runs counter to the reason for seeking
orthogonality in the first place.

\begin{figure}
    \centering
    \includegraphics[width=0.9 \columnwidth]{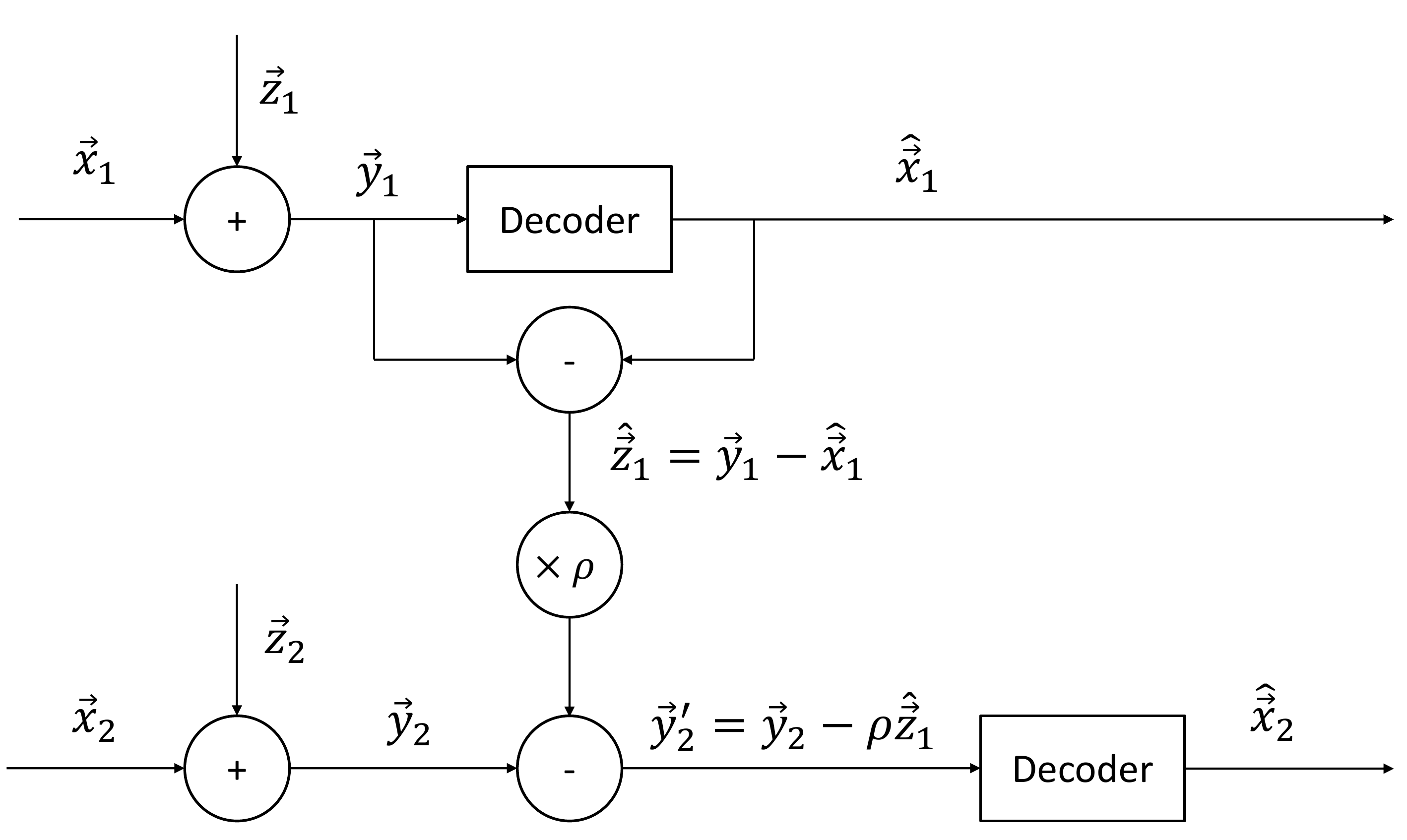}
    \caption{In noise recycling, a noise estimate is created from a lead channel by subtracting its modulated decoding from the received signal. That estimate is used to reduce noise on a channel subject to correlated noise prior to decoding.}
    \label{fig:system_model}
    \vspace{-0.2in}
\end{figure}

As detailed in Section \ref{sec:model}, we consider a particular
type of correlated noise model across orthogonal channels, Gauss-Markov
(GM)~\cite{anderson2012optimal} noise. The Gauss-Markov process has been
used to model progressive decorrelation of noise with growing
separation among channels \cite{NY07, AMM05, MG02, AG07} in time,
frequency, or both. Within that model, we introduce a novel approach that embraces noise correlation to significantly improve decoding performance while maintaining separate decoding over orthogonal channels.

In interference cancellation in multiple access channels, decoded
codewords are subtracted from the received signal to remove
interference \cite{cover2012elements, Gal85}. In contrast, in noise
recycling modulated decoded codewords are subtracted from the
received signal to recover noise estimates. That noise estimate is
a component of the noise in another as-yet undecoded orthogonal
channel due to correlation across channels. A proportion of the
estimate can, therefore, be subtracted from the received signal on
the orthogonal channel before decoding, reducing the latter's
effective noise.

Figure~\ref{fig:system_model} provides an illustration of the
technique. It shows two independent channel inputs, $(\vx_1,\vx_2)$
from potentially different codebooks, on orthogonal channels that
are corrupted by correlated real-valued additive symmetric Gaussian noise
$(\vZ_1,\vZ_2)$ with correlation $\rho$.
This results in correlated random real-valued channel outputs
$(\vY_1,\vY_2)= (\vx_1,\vx_2)+(\vZ_1,\vZ_2)$. For a particular realization of outputs, $(\vy_1,\vy_2)$,
on decoding the lead channel $\vy_1$ to $\vxhat_1$, the decoder estimates
the noise realization experienced on the lead channel by subtracting the decoded codeword from the received signal $\vzhat_1
= \vy_1-\vxhat_1$. The second receiver updates its channel output to
$\vy_2'=\vy_2-\rho\vzhat_1$, eliminating part of the additive noise
experienced on the second channel $\vz_2$, before decoding. This noise recycling results
in the second channel output being a less noisy version of the
channel input $\vx_2$, which in turn leads to improved decoding
performance.

We consider both the rate gain and Block Error Rate (BLER) improvements
yielded by noise recycling  vis-\`a-vis independently decoding the
channels. For rate gain, we provide
a proof of achievability with an ordering for the successive decoding
of orthogonal channels using noise recycling in Section \ref{subsec:proof}. We evaluate rate gains numerically, which improve both with correlation and with the number of orthogonal channels for a given correlation.

For BLER improvements, we illustrate that noise recycling can work
with any codes at any rates using any decoders on any channels, since
noise recycling only uses noise estimates. We consider two cases. The
first, discussed in Section \ref{sub:predetermined_order}, is similar
in spirit to the approach presented in Section \ref{sec:rate_gain}
to achieve rate gains where a low rate code is reliably decoded on
a lead channel, giving an accurate estimate of the noise on that
channel, whereupon a second, higher rate, code can be more reliably
decoded on an orthogonal channel subject to correlated noise.

The second case of BLER improvement, presented in Section
\ref{sub:same_rate}, does not pre-determine
which of the orthogonal channels is decoded first. Instead, the decoders of orthogonal channels are run in parallel, in effect racing each other. 
The first decoder to terminate provides the initial
noise estimate for noise recycling. While this approach is not
designed to provide rate gains, we show that it yields considerable
BLER improvements for short codes. With the increasing demand
for low-latency communications, high rate short codes are expected
to become more common.

\section{System Model}\label{sec:model}
\subsection{Definition, Notation and System Model}
Let $x,\vx,X, \vX$ denote a scalar, vector, random variable, and random vector, respectively. All vectors are row vectors. A linear block code is characterized by a code-length, $n$,  code-dimension, $k$, $\sbrace{n,k}$, and rate $R=k/n$. The binary field is denoted by $\gftwo$. Mutual information between $X,Y$ is denoted by $\mutualinf{X}{Y}$. We study an orthogonal channel system where $i\in\{1,\ldots,m\}$
messages, $\vu_i\in\gftwo^{k_i}$, are encoded into codewords
$\vc_i$. The codewords are modulated into $\vx_i$ and sent over
analog orthogonal channels subject to additive noise. Channel outputs
are $\vY_i=\vx_i+\vZ_i$, where $Z_{i,j}$ is the $j$-th element of
$\vZ_i$. For each $i$, we call the orthogonal channels $i$ and $i+1$
\textit{sequential orthogonal channels}. Noise is assumed to follow
a GM model where the $j$-th element of the $i$-th noise vector $\vZ_i$ is generated
in the following way: $Z_{i,j}=\rho Z_{i-1,j}+\Xi_{i,j}$ and the
innovation processes, $\cbrace{\Xi_{i,j}}_i$, are all mutually
independent and identically distributed
$\Xi_{i,j}\sim\cN\paren{0,\paren{1-\rho^2}\sigma^2}$, so that
$Z_{i,j}\sim\cN\paren{0,\sigma^2}$, for $\abs{\rho}<1$.
The rate of the $i$-th code is $R_i=k_i/n$,
and the total rate is $R=\sum_{i=1}^m R_i$. Given $(\vY_1,\ldots,\vY_m)$,
the goal is to estimate $(\vc_1,\ldots,\vc_m)$ using $m$ distinct decoders.

\section{Noise Recycling Rate Gain}\label{sec:rate_gain}
We first determine the rate region that can be achieved
by sending at a lower rate on one orthogonal channel, estimating the
realization of the noise in that channel, and using that knowledge
to reduce the impact of noise in other orthogonal channels.
\begin{theorem}
Assume a GM noise model with fixed $m$ orthogonal channels, each
with variance $\sigma^2$, and correlation $\rho$ between sequential
orthogonal channels. For a given average power constraint,
$\mathbb{E}\paren{X_i^2}\leq P$, and any correlation factor
$\abs{\rho}<1$, the following region is achievable:
\begin{align*}
R_1<C\paren{\frac{P}{\sigma^2}},\quad R_j<C\paren{\frac{P}{\paren{1-\rho^2}\sigma^2}}, \text{ for all } j>1,
\end{align*}
where $C\paren{P/\sigma^2}=0.5\log(1+P/\sigma^2)$ and so $P/\sigma^2$ is the the Signal
to Noise Ratio (SNR).
\end{theorem}

\subsection{Code construction for achievability}
\label{subsec:code}
\begin{figure*}
    \centering
    \includegraphics[width=1.9 \columnwidth]{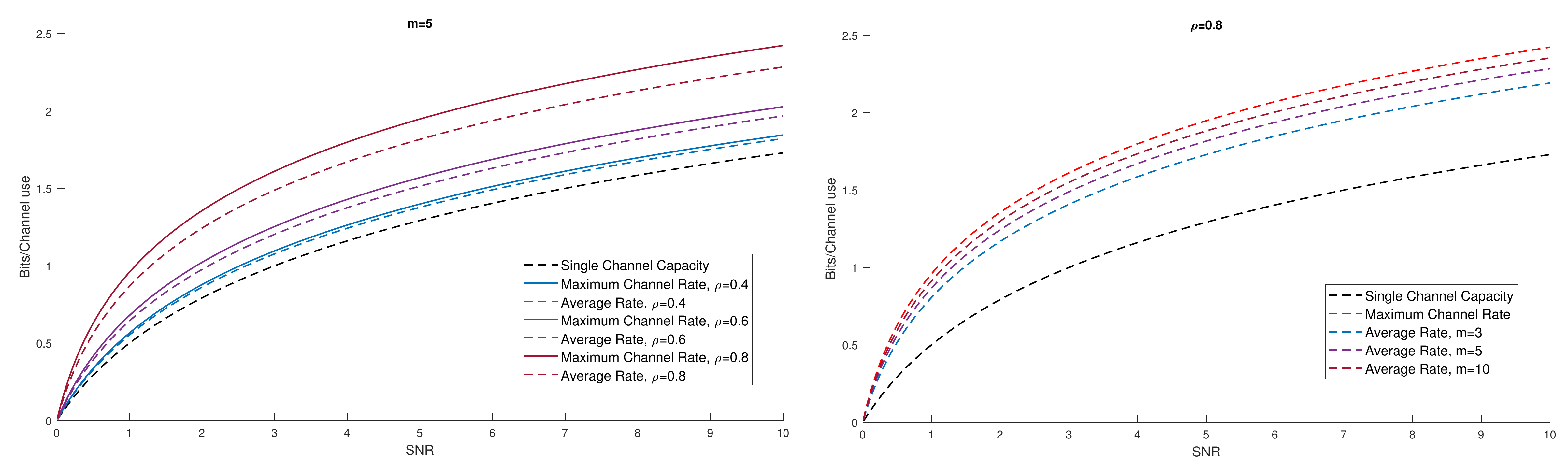}
    \caption{Achievable regime for the decoder of Section~\ref{sec:rate_gain}. Single Channel Capacity is  $C_1=C\paren{P/\sigma^2}=0.5\log\paren{1+P/\sigma^2}$, the capacity of an orthogonal channel that does not use recycled noise, as is the case for the first channel, $j=1$. Maximum Channel Rate is the rate of an orthogonal channel decoded with noise recycling $C_j=C\paren{P/(\paren{1-\rho^2}\sigma^2)}$, $j>1$. The average rate is the average rate per orthogonal channel, namely $(C_1+\paren{m-1}C_2)/m$.}
    \label{fig:capacity}
    \vspace{-0.18in}
\end{figure*}

To  establish constructively the theorem, our coding scheme carries
the flavor of~\cite[Chapter 9.1]{cover2012elements}. We create
$m$ independent random codebooks such that the $j$-th codebook
consists of $2^{nR_j}$ codewords independently drawn from
$\vX^j\paren{1},\ldots,\vX^j\paren{2^{nR_j}}\sim\cN\paren{0,P-\epsilon}$, where
$R_1<C\paren{P/\sigma^2},R_j<C\paren{P/\paren{1-\rho^2}\sigma^2}\forall
j\neq 1$, and a superscript $j$ indicates that a codeword was
chosen from the $j$-th codebook. For each channel, codebooks are
known to transmitter and receiver, but each channel need only know
its own codebook. The transmitters of the orthogonal channels send
$\vX^1\paren{i_1},\ldots,\vX^m\paren{i_m}$. The decoders operate sequentially from $j=1$ to $m$ as follows. With $\vZhat_0=0$ and, $\vZhat_{j-1}$ being the noise estimated from the decoding of the $j-1$-th channel, the $j$-th decoder subtracts
$\rho\vZhat_{j-1}$ from its channel output $\vY_j$, resulting in
$\vY_j'=\vY_j-\rho\vZhat_{j-1}$. It then identifies as the decoding the
unique codeword that is jointly-typical with $\vY_j'$ and satisfies
the power constraints. If
a codeword does not exist or is not unique, the $j$-th decoder
decodes in error. In the mathematical treatment, an error occurs
if any of the $m$ decoders is in error, but in simulations it errors are accounted for separately on each channel. Figure~\ref{fig:capacity}
illustrates the rate that can be gained when compared to the case where
channel decoders operate independently. It is evident that there is a
gap between the single-channel capacity and the average rate that
can be achieved by noise recycling decoders. In particular, there is a significant
rate-gain even when the number of orthogonal channels,
$m$, or the noise correlation, $|\rho|$, is low.

\subsection{Achievability proof}
\label{subsec:proof}
\begin{proof}
We prove the result using techniques redolent of those in~\cite[Chapter
9.1]{cover2012elements}. We bound from below the probability that the
jointly decoding $\vX_1,\ldots,\vX_m$ successfully. The event that
the first $j$ decodings are successful is denoted by
$\cC_j$. The event of a decoding failure in the
$j$-th decoder is denoted by $\cE_j$. Define the following events:
$E_{0,j}=\cbrace{n^{-1}\sum_{i}{\paren{X_i^j}^2}>P}$,
$E_{i,j}=\cbrace{\vX_j\paren{i},\vY_j'\text{ are jointly
}\epsilon\text{-typical}}$. Then
$=P\paren{\cC_m}=P\paren{\cC_1}\prod_{j=2}^m P\paren{\cC_j\mid \cC_{j-1}}$. From results for a single channel, we know that $P\paren{\cC_1}\geq 1-3\epsilon$ for $n$ sufficiently large. Without loss of generality, assume that the $j$-th transmitter sends the first codeword of the $j$-th codebook. We bound $P\paren{\cE_j\mid \cC_{j-1}}$ in a similar fashion to the single case:
\begin{equation*}
\begin{split}
    &P\paren{\cE_j\mid \cC_{j-1}}\leq \\
    &P\paren{E_{0,j}\mid \cC_{j-1}}+P\paren{E_{1,j}^c\mid \cC_{j-1}}+\sum_{i=2}^{2^{nR_j}}P\paren{E_{i,j}\mid \cC_{j-1}}\leq \\
    &\epsilon+\epsilon+\sum_{i=2}^{2^{nR_j}}P\paren{E_{i,j}\mid \cC_{j-1}}=2\epsilon+\sum_{i=2}^{2^{nR_j}}P\paren{E_{i,j}\mid \cC_{j-1}},
\end{split}
\end{equation*}
for $n$ sufficiently large, where the first inequality follows from the union bound, and the second inequality from the law of large numbers and joint typicality. We bound $P\paren{E_{i,j}\mid \cC_{j-1}},i>1$
\begin{equation*}
\begin{split}
&P\paren{E_{i,j}\mid \cC_{j-1}}=P\paren{E_{i,j}\mid X_{1},Y_{1},\ldots,X_{j-1},Y_{j-1}}=\\
&P\paren{E_{i,j}\mid X_{j-1},Y_{j-1}}\leq 2^{-n\paren{\mutualinf{X_j}{Y_j\mid X_{j-1}, Y_{j-1}}-R_j-3\epsilon}}
\end{split}
\end{equation*}using the Markov property, and
\begin{equation*}
\begin{aligned}
&\mutualinf{X_j}{Y_j\mid X_{j-1}, Y_{j-1}}\\
&=\mutualinf{X_j}{Y_j\mid X_{j-1}, X_{j-1}+Z_{j-1}}=\mutualinf{X_j}{Y_j\mid Z_{j-1}}\\
&=h\paren{X_j\mid Z_{j-1}}+h\paren{Y_j\mid Z_{j-1}}-h\paren{X_j,Y_j\mid Z_{j-1}}\\
&=h\paren{X_j}+h\paren{X_j+\rho Z_{j-1}+\Xi_j\mid Z_{j-1}}-\\
&\hspace{1cm}h\paren{X_j,X_j+\rho Z_{j-1}+\Xi_j\mid Z_{j-1}}\\
&=h\paren{X_j}+h\paren{x_j+\Xi_j}-h\paren{X_j,X_j+\Xi_j}\\
&=\mutualinf{X_j}{X_j+\Xi_j}=\mutualinf{X_j}{Y_j'}
\end{aligned}
\end{equation*}
using the fact that $\vX_j\perp\vZ_{j-1}$. Therefore, $P\paren{E_{i,j}\mid X_{j-1}, Y_{j-1}}\leq 2^{-n\paren{\mutualinf{X_j}{Y_j'}-R_j-3\epsilon}}$. Picking $R_j<\mutualinf{X_j}{Y_j'}-3\epsilon$ yields $P\paren{\cE_{i,j}\mid \cC_{j-1}}\leq 3\epsilon$. Ultimately, we get $P\paren{\cC_m}\geq \paren{1-3\epsilon}^m$ which concludes the proof as $\epsilon$ can be made arbitrarily small.
\end{proof}

\section{Noise Recycling BLER Improvement}\label{sec:reliability_gain}
In Section~\ref{sec:rate_gain} we determined the rate-gains available
from noise recycling through the use of random codebooks and joint
typicallity. Here we illustrate that BLER performance is enhanced
by noise recycling for existing codes and decoders.

We demonstrate the technique with CRC-Aided Polar
(CA-Polar) Codes~\cite{tal2011list,tal2015list}, which are Polar
codes~\cite{arikan2008channel,arikan2009channel} with an outer CRC
code and will be in 5G NR control channel
communications~\cite{3gppcapolar1}, and Random Linear Codes (RLCs)
which are known to be capacity achieving~\cite{gallager1973random},
but have been little investigated owing to the absence of efficient
decoders that can work at high rates until recent developments. For
decoders, we use the state-of-the-art CA-Polar-specific CRC-Aided
Successive Cancellation List decoder (CA-SCL)
~\cite{tal2011list,niu2012crc,tal2015list,balatsoukas2015llr,liang2016hardware}
as implemented in MATLAB's Communications Toolbox. We also use two
soft-information variants \cite{solomon2020,duffy2020} of the
recently introduced Guessing Random Additive Noise Decoder (GRAND)
\cite{duffy2018guessing, duffy2019capacity}, both of which can
decode any block code and are well suited to short, high-rate codes.
In simulations, we employ the GM channel described in
Section~\ref{sec:model} with Binary Phase Shift Keying (BPSK)
modulation.

\subsection{Predetermined decoding order approach}
\label{sub:predetermined_order}

We first consider a sequential decoding scheme akin to the one
described in Section~\ref{sec:rate_gain} where a lead channel is
decoded and a subsequent channel that has a higher rate is decoded
using noise recycled information. Block-errors are counted separately
on both the lead and subsequent channels as it is possible that the
subsequent channel decodes correctly, even if the lead channel is
in error.

In the first simulation, the lead channel encodes its data using
A CA-Polar code $\sbrace{256,170}$ with rate $R_1\approx 2/3$. The
second, orthogonal channel uses a higher rate CA-Polar code,
either $\sbrace{256,180}$ or $\sbrace{256,190}$ giving $R_2\approx
0.7$ or $0.74$ respectively. The noise correlation is set to
$\rho=0.5$ and is known to the second decoder. Both channels are
decoded with CA-SCL, with the second channel benefiting from noise recycling, where before decoding the noise estimate of the lead orthogonal channel is subtracted after multiplication by the noise correlation factor $\rho$.

\begin{figure}
    \centering
    \includegraphics[width=0.85 \columnwidth]{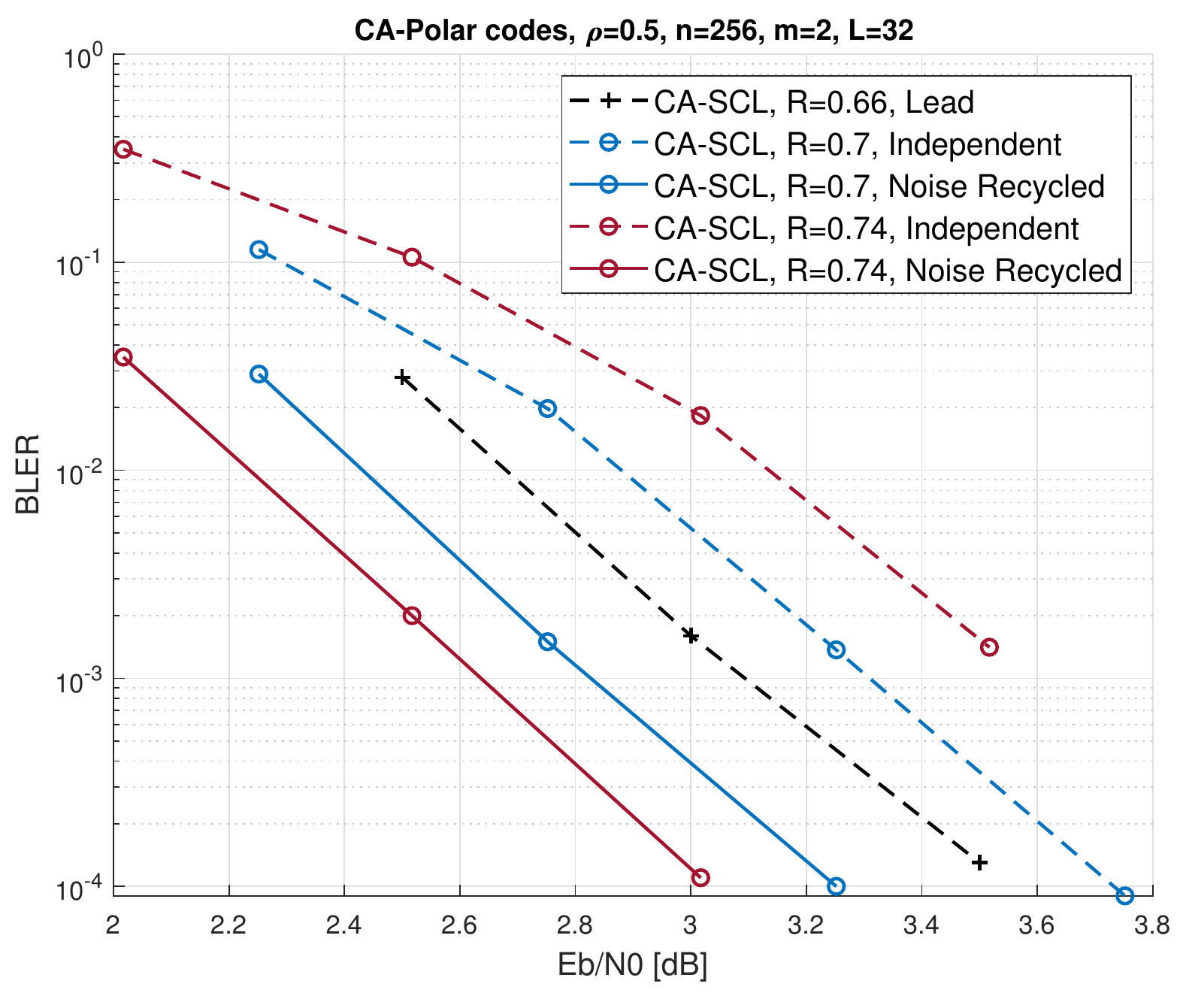}
    \caption{BLER vs. Eb/N0 for $256$ bit CA-Polar codes decoded
    with CA-SCL, uses a list size of $L=32$, with and without noise recycling. Dashed lines correspond to independent decoding,
    and solid lines to decoding after noise recycling. The
    lead orthogonal channel is encoded with a rate $2/3$ code. The
    second channel uses either a rate $0.7$ or $0.74$ code.}
    \label{fig:bler_diff_rate_m_2}
    \vspace{-0.15in}
\end{figure}

Figure~\ref{fig:bler_diff_rate_m_2} reports BLER vs Eb/N0. The black
dashed line corresponds to the lead channel, while the dashed blue
and red lines give the performance curves should noise recycling
not be used, corresponding to independent decoding of all channels.
As the second orthogonal channel runs at a higher rate than the lead channel, if decoded independently
the second channel would experience higher BLER than the lead channel. The solid blue
and red lines report the performance of the second decoder given
noise recycling. Despite using a higher rate code than the lead
channel, with noise recycling the second channel experiences better
BLER vs Eb/N0 performance. Notably, owing to the better Eb/N0 (i.e. the energy per {\bf information} bit used in the transmission) that
comes from running a higher rate code, the rate $0.74$ code provides better BLER than the rate $0.7$ code. For a commonly used target BLER of
$10^{-2}$, noise recycling results in $\approx 1$ dB gain for the
$\sbrace{256,190}$ code.

\begin{figure}
    \centering
    \includegraphics[width=0.85\columnwidth]{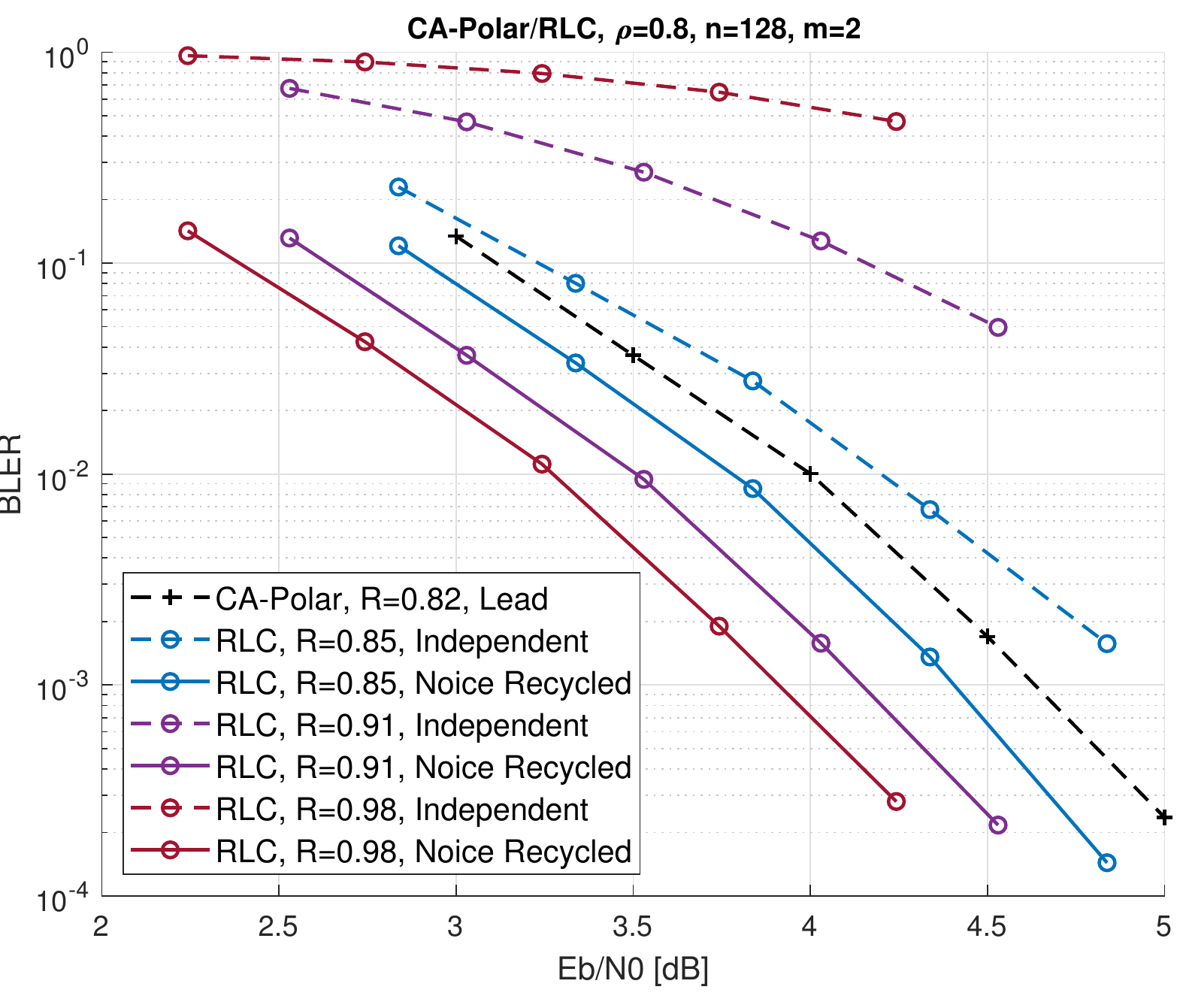}
    \caption{BLER vs. Eb/N0 for codes of length $n=128$
    decoded with ORBGRAND with and without noise recycling. Dashed lines correspond to independent decoding,
    and solid lines to decoding after noise recycling. Data on the
    lead orthogonal channel is encoded with a rate $0.82$ CA-Polar code. The
    second channel uses rate $0.85$, $0.91$ or $0.98$ RLCs.}
    \label{fig:bler_diff_rate_m_2_capolar_rlc}
    \vspace{-0.2in}
\end{figure}

Figure~\ref{fig:bler_diff_rate_m_2_capolar_rlc} reports an analogous
simulation, but where $\rho=0.8$, the lead channel's code is a
$\sbrace{128,105}$ CA-Polar code, $R_1=0.82$, and the second channel
is one of three RLCs ranging in rate from $0.85$ to $0.98$.  Both
channels are decoded with the recently proposed Ordered Reliability
Bits Guessing Random Additive Noise Decoding (ORBGRAND)~\cite{duffy2020}.
ORBGRAND is a soft detection decoder that has been reported to
provide more accurate decodings of CA-Polar codes than CA-SCL for
short codes. As with all the GRAND algorithms, it can decode any
code, making it viable for use with RLCs. A similar phenomenology
to the previous figure can be seen, where the impact of noise recycling
is even more dramatic, allowing the second channel code to
use reliably a much higher rate than the lead channel.

\begin{figure*}
    \centering
    \includegraphics[width=1.9 \columnwidth]{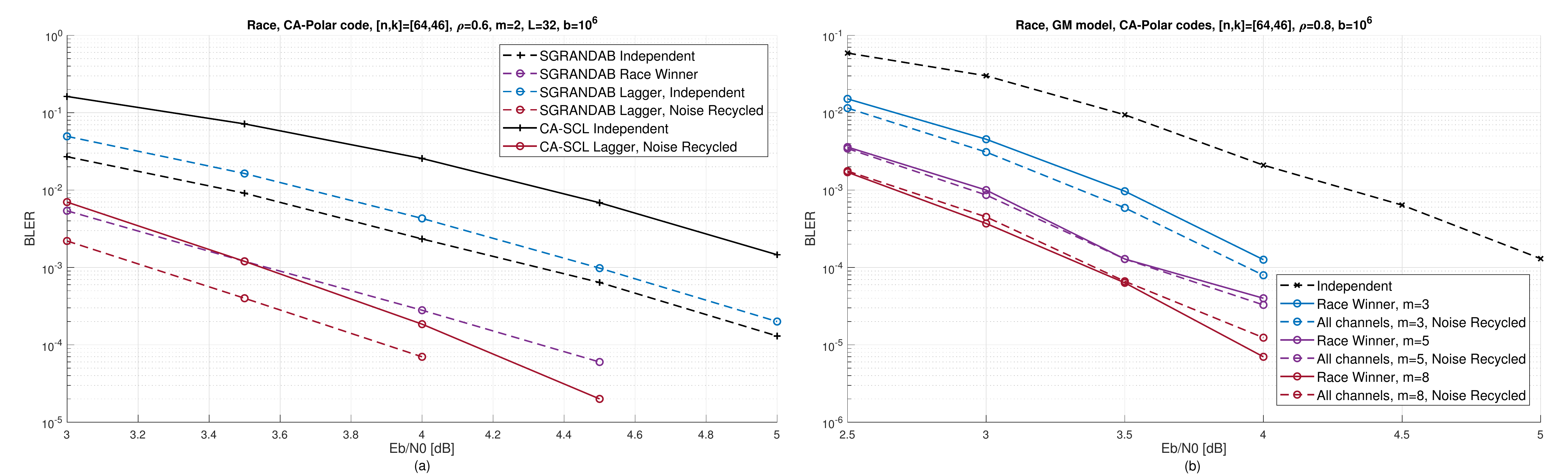}
    \caption{BLER vs Eb/N0 for $\sbrace{64,46}$ CA-Polar codes with racing. SGRANDAB, which uses an abandonment threshold of $b=10^6$, is used for the race. In (a), either SGRANDAB or CA-SCL, which uses a list size of $L=32$, are used to decode the laggers after noise recycling. In (b), all decoding is done using SGRANDAB.}
    \label{fig:bler_same_rate_m_mg}
    \vspace{-0.2in}
\end{figure*}

\subsection{Additional gains for short codes by racing}\label{sub:same_rate}

While previous sections identified rate and BLER improvements that
are available from running a pre-determined lead channel with a
lower rate code so that an accurate inference of a noise realization
could be obtained to aid the signal at a higher rate second channel,
here we consider an alternate design that leverages a significant
effect only observable with short codes. The principle behind noise recycling with racing is that all channels
use the same code-rate and orthogonal channels initially attempt
to decode their outputs contemporaneously. For certain types of
decoder, speed of decoding provides a measure of confidence in the
decoding accuracy and hence the precision of the noise estimate.
Thus once one decoder has identified a codeword, it is determined
to be the lead channel, other decoders cease their decodings, remove
the recycled noise estimate from the lead channel from their received
signal and decode. Noise recycling continues until all orthogonal
channel outputs have been decoded.

This decoding procedure is described in Algorithm~\ref{alg:same_rate}.
For example, suppose there are $m=3$ orthogonal channels. At the
first step, all decode in parallel. If decoder 2 is the
first to finish decoding, it is declared the winner of the race. The winner acts as lead channel
and provides an estimate $\vzhat_2$ to the decoders 1 and 3,
which repeat the process. Mixing-and-matching of decoders, even at different stages of the race, is still possible. This offers, for example, the possibility of using at the race phase a decoder that is highly accurate, but with potentially poor in worst-case runtime. As the race winner will terminate generally quickly, a decoder with uncertain termination time may not necessarily be deleterious in the race phase. Substituting a different deciding algorithm after noise recycling is then a possibility.

We demonstrate the race approach using a recently proposed technique,
Soft GRAND with ABanonment (SGRANDAB)~\cite{solomon2020}, that has the required feature. SGRANDAB aims to identify the noise that
corrupted a transmission from which the codeword can be inferred, rather than identifying the codeword directly. It does this by removing possible noise effects, from most likely to least likely as determined by soft information, from a received signal and querying whether what remains is in the codebook.
The first instance that results in success is a maximum likelihood decoding. If no codeword is found before a given number of codebook queries, SGRANDAB abandons decoding and reports an error. The channel that is decoded with the fewest codebook queries by
SGRAND wins the noise recycling race.

\begin{algorithm}
\caption{Noise recycling with racing}
\label{alg:same_rate}
\begin{flushleft}
        \textbf{Input:} $\vy_1,\ldots,\vy_m$\newline
        \textbf{Output:} $\vchat_1,\ldots,\vchat_m$
\end{flushleft}
\begin{algorithmic}[1]
\State Decode orthogonal channel outputs
\State $i\gets\text{ \# of decoder that won the decoding race}$
\State $\vchat_i\gets \;i\text{-th decoded codeword}$
\State $\vxhat_i\gets \;\text{encode of }\vchat_i$
\State $\vzhat_i\gets \vy_i-\vxhat_i$
\For{$j=1\rightarrow \max\cbrace{m-i,i-1}$}
\If{$i+j\leq m$}
\State $\sbrace{\vchat_{i+j},\vzhat_{i+j}}\gets\text{DecodeAndEst.}\paren{\vy_{i+j},\vzhat_{i+j-1}}$
\EndIf
\If{$i-j\geq 1$}
\State $\sbrace{\vchat_{i-j},\vzhat_{i-j}}\gets\text{DecodeAndEst.}\paren{\vy_{i-j},\vzhat_{i-j+1}}$
\EndIf
\EndFor
\State \Return $\vchat_1,\ldots,\vchat_m$
\end{algorithmic}
\begin{algorithmic}
\Procedure{DecodeAndEst.}{$\vy,\vzhat,j$}
\State $\vyhat\gets\vy-\rho\vzhat$
\State Decode orthogonal channel $j$ using $\vyhat$
\State $\vchat\gets\; \text{decoded codeword}$
\State $\vxhat\gets \;\text{encode of }\vchat$
\State $\vzhat\gets \vy-\vxhat$
\State \Return $\vchat,\vzhat$
\EndProcedure
\end{algorithmic}
\end{algorithm}

We simulated the noise recycling race in the presence of $m$ GM channels using a $\sbrace{64,46}$ CA-Polar code. We first consider the method on $m=2$ orthogonal channels with $\rho=0.6$, where the race, is done using SGRANDAB on all orthogonal channels, and the noise-recycled decoding is performed using either SGRANDAB or CA-SCL. Figure~\ref{fig:bler_same_rate_m_mg} (a) reports BLER performance. SGRANDAB outperforms CA-SCL without noise recycling, so it provides a better noise estimate following the decoding of the orthogonal channel that wins the race. The lagging channel benefits from noise recycling. It can continue to use SGRANDAB as during the race, or instead use CA-SCL. These results again shows a gain of more than 1 dB can be achieved, even for codes of the same rate, by racing   noise recycling. Figure~\ref{fig:bler_same_rate_m_mg} (b) reports the BLER of a decoder without noise recycling for $\rho=0.8$ and $m=3$, 5 or 8, the race winner with noise recycling, and all $m$ decoders after noise recycling, where all decoders use SGRANDAB. This shows a significant improvement in BLER for all values of $m$. For example there is a gain of about 1.7 dB for $m=8$ at a target BLER of $10^{-4}$. While this race advantage disappears as a consequence of averaging for long codes, it can be seen to provide a significant advantage for short codes.

\section{Conclusion and Discussion}\label{sec:conclusion}
We presented a novel way to recycle noise in orthogonal channels in order to improve communication performance for any combination of codes. The performance improvement is twofold, we proved its rate gain aspect and showed empirical evidence of its reliability improvement aspect. We analyzed orthogonal correlated channels, i.e. channels in which data that is sent on different channels is independent. A natural extension is considering the use of noise recycling in wireless communications, and the consequences of uncertainty in it. Noise recycling points to the benefit of correlation among orthogonal channels, opening an interesting vein of investigation where orthogonal channels, say in OFDM or TDMA, are chosen with a preference for noise correlation among them, with attendant effects in terms of rate and power allocation among orthogonal channels.

\bibliographystyle{IEEEtran}
\bibliography{bib}

\end{document}